\documentclass[12pt,article]{article}
\usepackage[T2A]{fontenc}
\usepackage[utf8]{inputenc}
\usepackage{amsthm,amsfonts,amssymb,amscd,cite}
\usepackage[english]{babel}
\usepackage[pdftex]{graphicx}
\usepackage{graphicx}
\usepackage[tbtags]{amsmath}
\usepackage{hyperref}

\DeclareSymbolFont{bletters}{OML}{cmm}{bx}{it}
\DeclareMathSymbol{\bla}{\mathord}{bletters}{'025}
\DeclareMathSymbol{\bmu}{\mathord}{bletters}{'026}
\DeclareMathSymbol{\bth}{\mathord}{bletters}{'022}
\DeclareMathSymbol{\bfI}{\mathord}{bletters}{"49}
\DeclareMathSymbol{\bdl}{\mathord}{bletters}{"0E}
\DeclareMathSymbol{\bDl}{\mathord}{bletters}{"001}

\def \bphi{\boldsymbol\phi}

\def \si{\sigma}
\def \la{\lambda}
\def \be{\beta}

\def \ta{\theta}
\def \dl{\delta}

\def \Dl{\Delta}

\def \CK{\mathcal M}
\def \CM{\mathcal M}
\def \CN{\mathcal N}
\def \CV{\mathcal V}

\def \CP{\mathcal P}

\def \CT{\mathcal T}
\def \BC{\mathbb{C}}
\def \BN{\mathbb{N}}
\def \BR{\mathbb{R}}

\def \BI{\mathbb{I}}

\begin{document}
\title{$${}$$\\
{\bf Correlation Functions as Nests of Self-Avoiding Paths}
}
\author{
{\bf Nikolay M. Bogoliubov$^{\star,\dagger}$, Cyril Malyshev$^{\star, \dagger}$}\\
$${}$$\\
$^{\star}${\it St.-Petersburg Department of Steklov Institute of Mathematics, RAS}\\
{\it Fontanka 27, St.-Petersburg,
RUSSIA}\\
$^{\dagger}${\it ITMO University} \\
{\it Kronverksky 49, St.-Petersburg, RUSSIA}
}

\date{}

\maketitle

\begin{abstract}
\noindent We discuss connection between the $XXZ$ Heisenberg spin chain in the limiting case of zero anisotropy and some aspects of enumerative combinatorics. The representation of the Bethe wave functions via the Schur functions allows to apply the theory of symmetric functions to calculation of the correlation functions. We provide a combinatorial derivation of the dynamical correlation functions of the projection operator in terms of nests of self-avoiding lattice paths.

\end{abstract}

\thispagestyle{empty}
\newpage

\section{Introduction}

The theory of {\it random walks}, being one of the classical directions of enumerative combinatorics \cite{stan}, was successfully applied in various fields: in the theory of quantum computations \cite{cw} and in the analysis of stock markets \cite{ef}, in biology \cite{ks} and in psychology \cite{ps}, in self-organized criticality \cite{sr} and in population processes \cite{er}.

The `Random walks problem' in theoretical physics was first introduced by M.~Fisher \cite{fish}. Fascinating
connections to other research fields, such as {\it Young diagrams}, and
the theory of {\it random matrices}, have been revealed one after
another \cite{2, 3, 4, 5, 7, 8}.

Some sections of enumerative combinatorics \cite{stan} and the theory of symmetric functions \cite{macd} has come to play an important role in the theory of integrable models \cite{fad, kbi} and especially in the studies of correlation functions \cite{bmnph, bmumn}.
The aim of this paper is to  represent correlation functions of $XX0$ spin chain as sums over nests of self-avoiding lattice paths. The interpretation of
correlation functions of bosonic integrable models in terms of random walks in multidimensional simplectical lattices was given in \cite{bogmasig, bogmalver}.

\begin{figure}[h]
\centering
\includegraphics[scale=1.0]{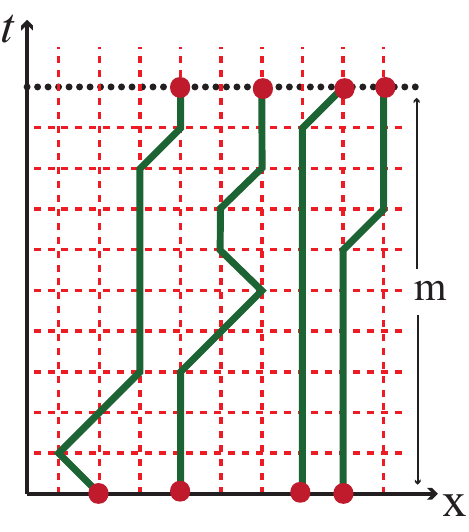}
\caption{Random turns walkers.}
\label{fig:f1}
\end{figure}

\begin{figure} [h]
\center\includegraphics{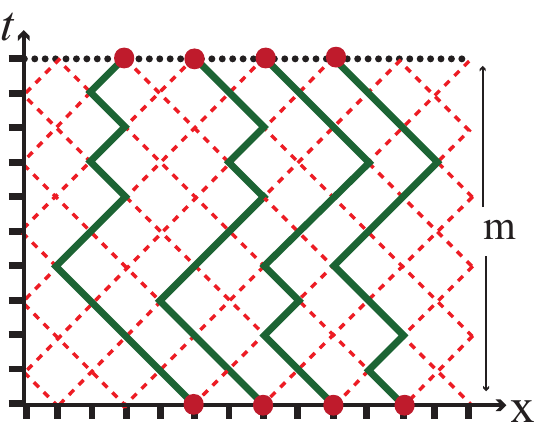}
\caption{Lock step walkers.}
\label{fig:f2}
\end{figure}

Two essentially different
types of vicious walkers may be distinguished in classification of \cite{fish}.
Suppose that there are $N$ walkers (particles) on a one-dimensional lattice.
For the {\it random turns model} at each tick of the
clock $dt$ only a {\underline {single}} randomly chosen walker moves one step to
the left or one step to the right while the rest are staying (Fig.~\ref{fig:f1}).
In the {\it lock step} version of the model at each tick of the
clock {\underline {each}} walker moves to the left or to the right lattice site with
equal probability (Fig.~\ref{fig:f2}).
Trajectories of random walkers can be viewed as directed lattice paths (i.e., the paths that cannot turn back)
which start at sites say on the line $x$ and finish after $m$ steps on sites on the line $t = m$.
Walkers are `vicious' so that two or more walkers are prohibited to
arrive at the same site simultaneously.

\begin{figure} [h]
\center\includegraphics{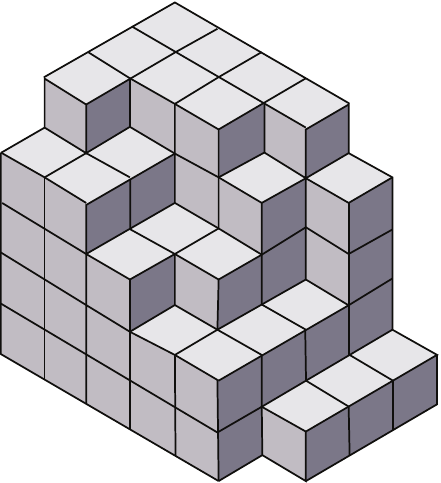}
\caption{Plane partition.}
\label{fig:f3}
\end{figure}

Random walks are closely related to {\it plane partitions} or three-dimensional Young diagrams. A plane partition is a two-dimensional array of nonnegative integers $n_{i, j}$ that are non-increasing both from left to right and from top to bottom: $n_{i, j}\geq n_{i, j+1}$ and $n_{i, j} \geq n_{i+1, j}$.
Plane partitions may be represented as a stack of $n_{i, j}$ unit cubes above the point $(i, j)$ (Fig.~\ref{fig:f3}).

Our paper is structured as follows. The current Section~1 is introductory. Section~2 is devoted to the {\it free fermion limit} of the $XXZ$ Heisenberg model. In Section~3 the correlation functions over zero particles ground state are discussed. In Section~4 the  combinatorial description of the thermal correlation functions is given. In Section~5 it is shown that the thermal correlation function of the projection operator
can be treated in terms of sums over nests of self-avoiding lattice paths of special type. An identity is obtained as a by-product which relates a trigonometric sum to an
integer equal to a number of self-avoiding lattice paths.
Discussion in Section~6 concludes the paper.

\section{$XXZ$ Heisenberg Spin Chain and Its Zero Anisotropy Limit}

The Heisenberg $XXZ$ model on the chain of $M+1$ sites is defined by the Hamiltonian
\begin{equation}
H_{\rm XXZ} = -\frac 12\sum_{k=0}^M
\bigr(\si_{k+1}^{-}\si_k^{+} +
\si_{k+1}^{+}\si_k^{-}+\frac\Dl
2\,(\si_{k+1}^z\si_k^z-\BI)
+ (\si_k^z-\BI)\bigl)\,,
\label{xxzham}
\end{equation}
where $\Dl\in\BR$ is the anisotropy parameter. The local spin operators $\si^\pm_k = \frac12 (\si^x_k\pm i\si^y_k)$ and $\si^z_k$ act nontrivially on $k^{\rm th}$ site and  obey the commutation rules:
\[
[\,\si^+_k, \si^-_l\,]\,=\,\dl_{k
l}\,\si^z_l\,,\quad
[\,\si^z_k,\si^\pm_l\,]\,=\,\pm 2\,\dl_{k
l}\,\si^{\pm}_l
\]
($\dl_{k l}$ is the
Kronecker symbol). Besides, $\BI$ acts in (\ref{xxzham}) as identity operator at $k^{\rm th}$ site. The spin operators act in the space ${\mathfrak H}_{M+1}$ spanned over the states $\bigotimes_{k=0}^M
\mid\!\! s\rangle_{k}$, where $\mid\!\!
s\rangle_{k}$ implies either spin ``up'',
$\mid\uparrow\rangle$, or spin ``down'',
$\mid\downarrow\rangle$, state at $k^{\rm th}$ site. The states $\mid\uparrow\rangle\equiv
\begin{pmatrix}
1 \\
0
\end{pmatrix}$ and $\mid\downarrow\rangle
\equiv
\begin{pmatrix}
0 \\
1
\end{pmatrix}$ provide a natural basis of the linear space ${\BC}^2$.
The state $|\!\!\Uparrow \rangle $ with all spins ``up'': $\mid\Uparrow\rangle
\equiv \bigotimes_{n=0}^M \mid \uparrow
\rangle_n$ is annihilated by the Hamiltonian (\ref{xxzham}):
\begin{equation}
H_{\rm XXZ}\mid \Uparrow \rangle=0\,. \label{fer}
\end{equation}
The Hamiltonian (\ref{xxzham}) commutes with the operator ${S}^z$ of the third component of the total spin:
\begin{equation}
\lbrack{H}_{\rm XXZ}, {S}^z]\,=\,0\,,\qquad
{S}^z\equiv \frac 12\sum_{k=0}^M\sigma_k^z\,.
\nonumber
\end{equation}

We shall consider the $XX$ Heisenberg model, which is the free fermion limit of the $XXZ$ Heisenberg spin chain. The Hamiltonian of the $XX$ spin chain arises as the
zero anisotropy limit $\Delta\to 0$ of the Hamiltonian (\ref{xxzham}):
\begin{equation}\label{xxham}
H_{\rm XX} =  \frac 12\, {\cal H}-\frac 12 \sum_{k=0}^{M}(\si_k^z - \BI)\,,
\end{equation}
where ${\cal H}$ is the ``hopping'' part:
\begin{equation}
{\cal H} \equiv - \sum_{k=0}^{M}
(\si_{k+1}^{-}\si_k^{+} +
\si_{k+1}^{+}\si_k^{-})\,. \label{ham}
\end{equation}
The system described by the Hamiltonian (\ref{xxham}) is of interest, for example, in the construction of the theory of
quantum computations \cite{vk3}.

Consider an arbitrary state on a chain. It can be characterized by the number $N$ of spins ``down'' and the number ${\CK}\equiv M-N+1$ of sites with spin ``up''.
The $N$-particle state-vectors  $\mid\!\Psi({\bf u}_N)\rangle$, i.e., the states with $N$ spins ``down'', are convenient to express by means of the Schur functions:
\begin{equation}
\mid\!\Psi({\textbf u}_N)\rangle =
\sum\limits_{\bla \subseteq \{{\CK}^N\}}
S_\bla ({\textbf u}^{2}_N) \begin{pmatrix}
\prod\limits_{k=1}^N \si_{\mu_k}^{-}\end{pmatrix} \mid
\Uparrow \rangle\,. \label{conwf1}
\end{equation}
The sites with spin
``down'' states are labeled by the
coordinates $\mu_i$, $1\leq i\leq N$. These coordinates constitute a strictly decreasing partition ${\bmu}=(\mu_1, \mu_2,\,\dots\,, \mu_N)$, where the numbers ${\mu}_i$, called {\it parts}, respect the inequality $M\geq \mu_1> \mu_2 > \, \dots\,>
\mu_N \geq 0$. The relation
$\la_j=\mu_j-N+j$, where $1\le j\le N$, connects the parts of $\bla$ to those of $\bmu$. Therefore, we can write: $\bla =\bmu - {\bdl}_N$, where ${\bdl}_N$ is the strict
partition
\begin{equation}
{\bdl}_N \equiv (N-1, N-2, \dots, 1, 0)\,.
\label{strict}
\end{equation}
Besides, the bold notations like ${\bf u}^2_N\equiv (u^2_1, u^2_2, \dots , u^2_N)$ imply sets of arbitrary complex numbers. The summation in (\ref{conwf1}) is over all partitions
$\bla$ with parts satisfying ${\CK} \geq \la_1\geq \la_2\geq \dots\geq
\la_N\geq 0$.

\begin{figure} [h]
\center\includegraphics {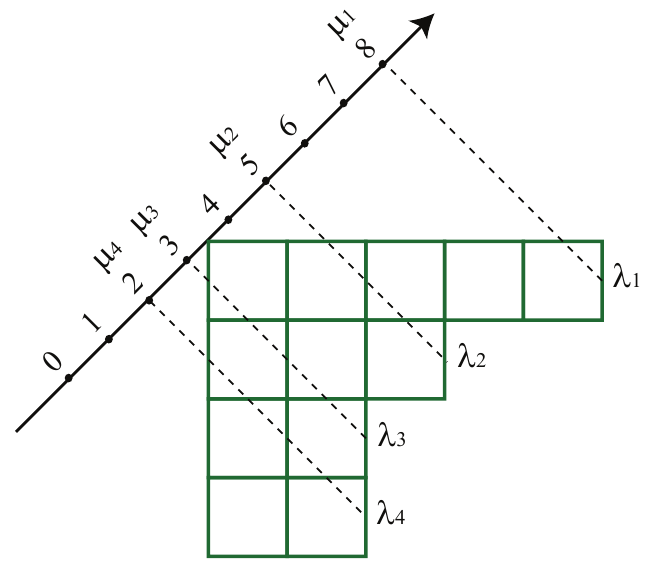}
\caption{Relation of the spin ``down'' coordinates $\bmu=(8, 5, 3, 2)$ and partition $\bla=(5, 3, 2, 2)$ for $M=8$, $N=4$.}
\label{fig:f4}
\end{figure}

The {\it Schur functions}
$S_\bla$ are defined
by the Jacobi-Trudi relation:
\begin{equation}
S_{\bla} ({\textbf x}_N)\,\equiv\,
\displaystyle{ S_{\bla} (x_1, x_2, \dots , x_N)\,\equiv\, \frac{\det(x_j^{\la_k+N-k})_{1\leq
j, k \leq N}}{\CV({\textbf x}_N)}}\,,
\label{sch}
\end{equation}
in which $\CV ({\textbf x}_N)$  is the Vandermonde determinant
\begin{equation} \CV ({\textbf x}_N)
\equiv
\det(x_j^{N-k})_{1\leq j, k\leq N}
\,=\,
\prod_{1 \leq m< l \leq N}(x_l-x_m)\,.
\label{spxx1}
\end{equation}
The conjugated state-vectors are given by
\begin{equation}\label{conj}
\langle \Psi({\bf v}_N)\mid\,=\,\sum\limits_{\bla \subseteq \{{\CK}^N\}}
\langle \Uparrow \mid \begin{pmatrix}
\prod\limits_{k=1}^N \si_{\mu_k}^{+}\end{pmatrix} S_\bla ({\textbf v}^{-2}_N)\,.
\end{equation}
There is a natural correspondence between the coordinates of the spin
``down'' states $\bmu$ and the partition $\bla$ expressed by the Young diagram (see Fig.~\ref{fig:f4}).

Assume that the periodic boundary conditions are imposed: $\si^{\#}_{k+(M+1)}=\si^{\#}_k$. If the
parameters $u^2_j\equiv
e^{i\ta_j}$ ($1\le j\le N$) satisfy
the Bethe equations,
\begin{equation}
e^{i (M+1)\ta_j}=(-1)^{N-1}\,, \quad 1\le j \le N \,,
\label{betheexp}
\end{equation}
then the state-vectors (\ref{conwf1})
become the eigen-vectors of the Hamiltonian (\ref {xxham}) \cite{kbi}:
\begin{equation}\label{egv}
{H}_{\rm XX}
\mid\!\Psi ({\bth }_N)\rangle\,=\, E_N({\bth }_N)
\mid\!\Psi ({\bth }_N)\rangle\,.
\end{equation}
The solutions $\theta_j$ to the Bethe equations (\ref{betheexp}) can be parameterized so that
\begin{equation}
\theta_j = \frac{2\pi
}{M+1}\begin{pmatrix}\displaystyle{
I_j - \frac{N-1}{2}} \end{pmatrix}\,,
\quad 1\le j \le N\,, \label{besol}
\end{equation}
where $I_j$ are integers or half-integers depending on whether $N$ is odd or even.

The eigen-energies in (\ref{egv})
are equal to
\begin{equation}
E_N({\bth}_N)\,=\,N - \sum_{j=1}^N\cos\ta_j\,=\,N -
\sum_{j=1}^N\cos\begin{pmatrix} \displaystyle{\frac{2\pi }{M+1}
\Bigl(I_j-\frac{N-1}{2}\Bigr)}\end{pmatrix}.
\label{egen}
\end{equation}
The \textit{ground state} of the model is the eigen-state that corresponds to the lowest eigen-energy $E_N(\bth^{\,\rm g}_N)$. It is determined by the solution to the Bethe equations (\ref{besol}) at $I_j=N-j$:
\begin{equation}\label{grstxx}
\theta^{\,\rm g}_j \equiv \frac{2\pi
}{M+1}\begin{pmatrix} \displaystyle{\frac{N+1}{2} -j} \end{pmatrix}\,, \quad
1\le j \le N\,,
\end{equation}
and is equal to
\[
E_N(\bth^{\,\rm g}_N)\, = \,N-\,\frac{
\,\sin\frac{\pi N}{M+1}}{\sin\frac{\pi}{M+1}}\,.
\]

In this paper we shall be restricted with the case of finite size of the system in order to consider the dynamical correlation function called the {\textit {persistence of ferromagnetic string}} and related to the projection operator $\bar\varPi_{n}$ that forbids spin ``down'' states on $n$ consecutive  sites of the chain \cite{bmnph}:
\begin{equation}
\CT ({\bth}^{\,\rm g}_N, n,
t)\,\equiv\,\frac{\langle
\Psi ({\bth}^{\,\rm g}_N)\mid
\bar\varPi_{n}\,e^{-t {H}_{\rm {XX}}}\,\bar\varPi_{n}
\mid\!\Psi ({\bth}^{\,\rm g}_N)\rangle
}{\langle \Psi ({\bth}^{\,\rm g}_N)\mid
e^{- t {H}_{\rm {XX}}} \mid\!\Psi ({\bth}^{\,\rm
g}_N)\rangle}\,,\qquad \bar\varPi_{n} \equiv \prod\limits_{j=0}^{n-1}\, \frac{\BI + \si^z_j}{2}\,, \label{ratbe0}
\end{equation}
where $t\in\BC$. We assume that
$\bar\varPi_{0}$ is the identity operator so that $\CT
({\bth}^{\,\rm g}_N, 0, \be)=1$.

\section{Correlations Over Zero Particles Ground State}

First, we shall consider the simplest one-particle correlation function
\begin{equation}
G\left(j, m | t \right)\equiv \langle \Uparrow \mid \si_j^{+} e^{-\frac t2\cal H}\si_m^{-} \mid \Uparrow \rangle\,,  \label{cf}
\end{equation}
where $\cal H$ is the Hamiltonian (\ref{ham}), which can be re-expressed through, so-called, {\it hopping matrix} $\bDl \equiv (\Dl_{n m})_{0 \le n, m\le M}$ \cite{8, bmtmf}. In the problem of vicious walkers it is more appropriate to use ${\cal H}$ expressed as follows:
\begin{equation}
{\cal H} = -\sum_{n, m =0}^M
\Dl_{n m} \si_n^{-} \si_m^{+}\,, \qquad \Dl_{n m} \equiv  \dl_{|n-m|, 1} + \dl_{|n-m|, M}\,.
\label{hamr}
\end{equation}
Differentiating $G\left(j, m | t\right)$ with respect to $t$ and
applying the commutation relation
\begin{equation}
\lbrack {\cal H},\si_m^{-}]= - \sum_{n=0}^M \Dl_{n m}\si_n^{-} \si_m^z\,, \label{cr}
\end{equation}
we obtain, \cite{8, bmtmf}, the difference--differential
equation:
\begin{equation}
\frac d{dt}\,G \left(j, m | t\right) = \frac 12\bigl(G \left(j, m-1 | t\right) + G \left(j, m+1 | t\right)\bigr)\,.
\label{eseq}
\end{equation}
Equation (\ref{eseq}) is supplied  at fixed $j$ with the periodicity requirement $G (j, m+M+1 | t )=G (j, m | t )$. An analogous requirement $G (j+M+1, m | t )=G (j, m | t )$ is valid for fixed $m$ as well. Besides, the ``initial condition'' is given by $G\left(j, m | 0\right)=\dl_{j m}$.

The correlator  $G\left(j, m | t\right)$ (\ref{cf}) may be considered as the exponential
generating function of random walks. Indeed, let us
introduce the notation $\mathcal D^K_{\lambda}$ for the operator
of differentiation of $K$-th order with respect to~$\lambda$ at
the point $\lambda=0$. Let us put the correlation
function (\ref{cf}) as a series in powers of $t$:
\begin{equation}\label{avv1}
G (j, m | t ) = \sum_{K=0}^{\infty}
\frac{(t/2)^K}{K!}\,\mathfrak{G} (j, m | K)\,.
\end{equation}
Acting by $\mathcal D^K_{t/2}$ on $G (j, m | t )$ (\ref{cf}) one obtains the coefficients $\mathfrak{G} (j, m | K)$ (\ref{avv1}) which are reduced
to the entries of the $K^{\rm th}$ power of $\bDl$
(\ref{hamr}):
\begin{eqnarray}
\mathfrak{G} (j, m | K) = \mathcal D^K_{t/2}\,G (j, m | t )
\nonumber \\ [0.2cm]
 = \langle \Uparrow \mid \si_j^{+} (-\,{\cal H})^K \si_m^{-} \mid
\Uparrow \rangle = \bigl({\bold\Delta}^K \bigr)_{j m}\,.\label{avv}
\end{eqnarray}
Indeed, applying the commutation relation (\ref{cr}), one obtains:
\begin{equation}
(-{\cal
H})^K \si_m^{-} \mid \Uparrow \rangle = \sum_{n=0}^M
\bigl({\bold\Delta}^K \bigr)_{n m}
\si_{n}^{-}\mid \Uparrow
\rangle \,. \label{lp}
\end{equation}
Equation (\ref{lp}) may be interpreted in the following way. Position of the walker on the chain is labelled by the spin ``down''
state, while the spin ``up'' states correspond to empty sites.
Each matrix ${\bold\Delta}$ in the product (\ref{lp}) corresponds to a transition between two neighboring sites. The
relation (\ref{lp}) enables to enumerate all admissible paths of the
walker starting from the $m^{\rm {th}}$ site. The state $\langle \Uparrow \mid \si_j^{+}$ acting on (\ref{lp}) from left allows to fix the ending point of the paths
because of the orthogonality of the spin states, and Eq.~(\ref{avv}) thus arises.

Let $|P_K(m \rightarrow j)|$ to denote the number of paths between the $m^{\rm {th}}$ and $j^{\rm {th}}$ sites. It is clear that $\mathfrak{G} (j, m | K) \equiv |P_K(m \rightarrow j)| =\bigl({\bold\Delta}^K \bigr)_{j m}$, and the generating function
$G (j, m | t )$ includes processes with all possible numbers of steps. It follows from Eq.~(\ref{avv}) that $\mathfrak{G}(j, m | K)$ satisfies the equation:
\begin{equation}\label{dmcf}
\mathfrak{G}(j, m | K+1)=\mathfrak{G}(j, m-1 | K)+ \mathfrak{G}(j, m+1 | K)\,,
\end{equation}
with the ``initial'' condition $\mathfrak{G}(j, m | 0)=\delta_{j m}$.

In theory of lattice paths the following generating function is usually used:
\begin{equation}\label{gfp}
F(j, m | z)=\sum\limits_{K=0}^\infty z^K \langle \Uparrow \mid \si_j^{+} (-{\cal H})^K \si_m^{-}\mid \Uparrow \rangle\equiv \sum\limits_{K=0}^\infty z^K \mathfrak{G}(j, m | K)\,,
\end{equation}
which is the Laplace transform of the exponential generating function (\ref{cf}):
\begin{equation}
\label{ltegf}
\int\limits_0^\infty G(j, m | t)\,
e^{-\frac{t}{z}} dt = z \langle \Uparrow \mid \si_j^{+} \begin{pmatrix} \displaystyle{ \frac{1}{1+z{\cal H}}}\end{pmatrix} \si_m^{-}\mid\Uparrow \rangle = z F(j, m | z)\,,\quad \Re z> 0\,.
\end{equation}

Consider now the multi-particle correlation function
\begin{equation}
G({\bf j}; {\bf l} | t)=\langle \Uparrow \mid \begin{pmatrix}
\prod\limits_{n=1}^{N}
\si_{j_n}^{+}
\end{pmatrix} e^{- \frac t2 \cal H} \begin{pmatrix}
\prod\limits_{k=1}^{N}
\si_{l_k}^{-}
\end{pmatrix}\mid \Uparrow \rangle\,,
\label{mpcf}
\end{equation}
which is parametrized by multi-indices ${\bf j}\equiv(j_1, j_2, \dots, j_N)$ and ${\bf l}\equiv(l_1, l_2, \dots, l_N)$.
This correlator is the generation function of $N$ random turns vicious
walkers (see Fig.~\ref{fig:f1}).
Really, let $|P_K ({\bf j} \rightarrow\,{\bf l})|$ be the number of $K$-edge paths traced by $N$ vicious walkers in the random turns model.
The commutation relation
\begin{equation}
[ {\cal H},\si_{l_1}^{-} \si_{l_2}^{-} \dots \si_{l_N}^{-}] = \sum_{k=1}^N \si_{l_1}^{-} \dots \si_{l_{k-1}}^{-} [{\cal
H},\si_{l_k}^{-}] \si_{l_{k+1}}^{-} \dots\si_{l_N}^{-}  \label{comcom}
\end{equation}
enables us to see that the average
\begin{equation}\label{qanal13}
\mathfrak{G}({\bf j}; {\bf l} | K) \equiv \mathcal D^K_{t/2}\,G ({\bf j}; {\bf l} | t) = \langle \Uparrow \mid \begin{pmatrix}
\prod\limits_{n=1}^{N}
\si_{j_n}^{+}
\end{pmatrix} (-\,{\cal H})^K \begin{pmatrix}
\prod\limits_{k=1}^{N}
\si_{l_k}^{-}
\end{pmatrix} \mid \Uparrow \rangle
\end{equation}
is equal to the number $|P_K ({\bf j} \rightarrow\,{\bf l})|$ of configurations of $N$ random turns walkers being initially located on the sites $l_1>l_2> \dots > l_N$ and arrived after $K$
steps at the positions $j_1>j_2> \dots > j_N$:
\begin{equation}
\label{qanal27}
|P_K ({\bf j} \rightarrow\,{\bf l})| = \mathfrak{G}({\bf j}; {\bf l} | K)\,.
\end{equation}
The condition that vicious walkers do not touch each other up to $N$ steps is guaranteed by the property of the Pauli matrices $(\sigma_k^{\pm })^2=0$.

Differentiating (\ref{mpcf}) by $t$ and applying (\ref{comcom}) we
obtain for fixed ${\bf j}$ the equation
\begin{eqnarray}
\frac d{dt}\,G ({\bf j}; {\bf l} | t) = \frac 12 \sum_{k=1}^N\bigl(G({\bf j}; l_1, l_2, \dots , l_k+1, \dots, l_N | t)&&  \nonumber\\
+ \,G({\bf j}; l_1, l_2, \dots, l_k-1, \dots, l_N | t)\bigr) &&
\label{phem}
\end{eqnarray}
(and a similar one for fixed
${\bf l}$). The non-intersection condition means that $G ({\bf j}; {\bf l} | t)=0$ if $l_k=l_p$ (or $j_k=j_p$) for any $1\leq
k, p\leq N$. The ``initial condition'' is: $G({\bf j}; {\bf l} | 0)=\prod_{m=1}^N \dl_{j_m, l_m}$.

The generating function $G({\bf j}; {\bf l} | t)$ respecting Eq.~(\ref{phem}) is given by the following

\vskip0.3cm \noindent {\large\sl Proposition.\,} \textit{Solution to Eq.}~(\ref{phem}) \textit{takes the form:}
\begin{equation}
G({\bf j}; {\bf l} | t)\,=\,
\displaystyle{\frac{e^{t N}}{(M+1)^N}
\sum\limits_{\{{\bphi}_N\}}
e^{- t E_N({\bphi}_N)}}|{\CV} (e^{i {\bphi}_N})|^2\,
S_{{\bla^L}}(e^{i {\bphi}_N})
S_{{\bla^R}}(e^{-i {\bphi}_N})\,,
\label{ratbe7}
\end{equation}
\textit{where the strict partitions ${\bf j}$ and ${\bf l}$ and the  partitions ${\bla^L}$ and ${\bla^R}$ are related:
${\bla^L}={\bf j}- {\bdl}_N$, ${\bla^R}={\bf l}- {\bdl}_N$.  The eigen-energy $E_N({\bphi}_N)$ is defined by} (\ref{egen}),
\textit{$e^{\pm i {\bphi}}\equiv (e^{\pm i\phi_1}, e^{\pm i\phi_2}, \dots, e^{\pm i\phi_N})$, and ${\CV} (e^{i {\bphi}_N})$ is defined by} (\ref{spxx1}).

\vskip0.3cm \noindent {\large\sl Proof.\,} It is easy to verify that the solution of (\ref
{phem}) is given by
\begin{equation}
G ({\bf j}; {\bf l} | t)\,=\,\det \bigl( G(j_r, l_s | t)\bigr)_{1\le r, s \le N}\,,  \label{det}
\end{equation}
where $G(j, l | t)$ is the one-particle generating function (\ref{cf})
satisfying Eq.~(\ref{eseq}).
The solution (\ref{det}) may be expressed in the form:
\begin{equation}
G({\bf j}; {\bf l} | t)\,=\,\displaystyle{
\frac{1}{(M+1)^N\,N !} \sum\limits_{s_1, \dots , s_N = 0}^M
e^{t\sum\limits_{m=1}^N \cos {\phi}_{{s}_m}}\,\det \bigl(
e^{i(j_r-l_s)\phi_{s_r}}\bigr)_{1\le
r, s \le N}}\,,
\label{ratbe6}
\end{equation}
where the parametrization is the same as in (\ref{ratbe7}).
The antisymmetry of the summand with respect to permutations of $\phi_1, \ldots , \phi_N$ enables to transform $\det \bigl(e^{i(j_r-l_s) {\phi}_{{s}_r}} \bigr)_{1\le r, s \le N}$ in (\ref{ratbe6}) into the product
of $\det \bigl(
e^{ij_r {\phi}_{{s}_r}}\bigr)_{1\le r, s \le N}$ and $\det \bigl(
e^{-il_s {\phi}_{{s}_r}} \bigr)_{1\le r, s \le N}$. So right-hand side of (\ref{ratbe6}) is expressed in terms of the Schur functions (\ref{sch}), and the representation (\ref{ratbe7}) is thus valid. It is clear that Eq.~(\ref{ratbe7}) at $N=1$ gives the solution to (\ref{eseq}). $\Box$

\vskip0.3cm \noindent {\large\sl Corollary.\,} From Eqs. (\ref{qanal13}) and (\ref{ratbe7}) we obtain that $\mathfrak{G}({\bf j}; {\bf l} | K)$ (\ref{qanal13}) is represented as the trigonometric sum:
\begin{eqnarray}
\mathfrak{G}({\bf j}; {\bf l} | K)\,=\,
\displaystyle{\frac{1}{(M+1)^N}}
\sum\limits_{\{{\bphi}_N\}}
\Bigr(2\sum\limits_{m=1}^N \cos {\phi}_{m}\Bigl)^K \nonumber \\
\times\,|{\CV} (e^{i {\bphi}_N})|^2\,
S_{{\bla^L}}(e^{i {\bphi}_N})
S_{{\bla^R}}(e^{-i {\bphi}_N})\,,
\label{ratbeco}
\end{eqnarray}
which takes according to (\ref{qanal27}) the integer value $|P_K ({\bf j} \rightarrow\,{\bf l})|$. $\Box$

In the particular case when ${\bf j} = {\bf l} = \bdl_N$, where $\bdl_N$ is defined by  (\ref{strict}), Eq.~(\ref{ratbeco})
is specified as
\begin{equation}\label{ratbe66}
  \mathfrak{G}({\bdl}_N; {\bdl}_N | K)=\displaystyle{\frac{1}{(M+1)^N}}
\sum\limits_{\{{\bphi}_N\}}
\Bigr(2\sum\limits_{m=1}^N \cos {\phi}_{m}\Bigl)^K |{\CV} (e^{i {\bphi}_N})|^2\,,
\end{equation}
since ${\bla^L} = {\bla^R} = (0, 0, \ldots, 0)$ and so \[S_{(0, 0, \ldots, 0)}(e^{i {\bphi}_N})=S_{(0, 0, \ldots, 0)}(e^{-i {\bphi}_N})=1\,.\]
In the thermodynamic limit the sum (\ref{ratbe66}) becomes the Gross-Witten partition function \cite{gross} and expresses, as well, the distribution of the length of the longest increasing subsequence of random permutations  \cite{joh}.

\section{Lattice Paths Interpretations of the Determinantal Representations}

In this section an algebraic approach to the calculation of the correlation functions $\CT ({\bth}^{\,\rm g}_N, n, t)$ (\ref{ratbe0}) is developed. The approach is based on  the {\it Cauchy-Binet formula}
for the Schur functions: \begin{equation}
\mathcal{P}_{L-n}(\textbf{y},
\textbf{x})\equiv\sum_{\bla\backslash {\bf n}\subseteq
{\{(L-n)^N\}}} S_{\bla}
(\textbf{x}) S_{\bla} (\textbf{y})\,=\,\begin{pmatrix}
\displaystyle{\prod_{l=1}^N y_l^n x_l^n}\end{pmatrix} \frac{\det_{N\times N} T(\textbf{x}, \textbf{y}) }{\CV_N(\textbf{x})\CV_N(\textbf{y})}\,,
\label{scschr}
\end{equation}
where $\sum_{\bla\backslash {\bf n}\subseteq
{\{(L-n)^N\}}}$ implies summation over all non-strict
partitions $\bla$ with the parts satisfying the inequality: \[
{L}-n\ge \la_1 -n \ge \la_2 -n\ge \dots \ge \la_N -n \ge 0\,.\]
The entries $(T_{k j}(\textbf{x}, \textbf{y}))_{1\le k, j\le N}$ of the $N\times N$ matrix $T(\textbf{x}, \textbf{y})$ in right-hand side of (\ref{scschr}) are expressed as
\begin{equation}\label{tt}
T_{k j}=\frac{1-(x_k y_j)^{L-n+N}}{1-x_k y_j}\,.
\end{equation}

The scalar products of the state-vectors, as well as the correlation functions, are connected with the generating functions of boxed plane partitions \cite{bmumn}. To study the asymptotical behaviour of the introduced correlation functions, we need the Cauchy-Binet relation (\ref{scschr}) taken in the $q$-parameterization \[\textbf{y}= \textbf{q} \equiv (q, q^2, \ldots, q^N)\,,\qquad \textbf{x}=\textbf{q}/q\equiv(1, q, \ldots, q^{N-1})\,.\] Letting $L={\CK}$, we obtain from (\ref{scschr}):
\begin{eqnarray}
&&\sum_{\bla\backslash {\bf n} \subseteq
\{(\CK-n)^N\}}S_{\bla} (\textbf{q}) S_{\bla}
(\textbf{q}/q)
\nonumber \\
&& = \frac{q^{nN^2}}{
\CV_N(\textbf{q})
\CV_N(\textbf{q}/q)}\det \begin{pmatrix} \displaystyle{ \frac{1-q^{(M+1-n)(j+k-1)}}{1-q^{j+k-1}}}
\end{pmatrix}_{1\leq j, k \leq N}
\nonumber \\ [0.2cm]
&& =q^{nN^2}\,q^{\frac {N({\CK}-n)}2 (1-{\CK}+n)}
\,\det \left(\displaystyle{
\begin{bmatrix} 2N+i-1 \\ N+j-1\end{bmatrix}}
  \right)_{1\leq i, j\leq {\CK}-n}
  \label{scqschr3}\\ [0.2cm]
&& =q^{nN^2} Z_q(N,N,\CK-n)
\label{scqschr}\,.
\end{eqnarray}
The entries in (\ref{scqschr3}) are the $q$-{\it binomial coefficients} defined as
\begin{equation}\label{qbc}
  \begin{bmatrix}R\\r\end{bmatrix}\, \equiv\,
\,\frac{[R]!}{[r]!\,[R-r]!}\,,\qquad [n]\,\equiv\,\frac{1-q^n}{1-q}\,.
\end{equation}
Besides, $Z_q (N, N, {\CK}-n)$ in (\ref{scqschr}) is
the MacMahon generating function of  plane partitions in the box ${\cal B}(N, N, \CK-n)$ of size $N\times N\times (\CK-n)$:
\begin{equation}\label{scpp}
Z_q (N,N,\CK-n)=\prod_{k=1}^{N} \prod_{j=1}^N
\frac{1-q^{\CK-n+j+k-1}}{1-q^{j+k-1}}\,.
\end{equation}
The number of plane partitions in ${\cal B}(N, N, \CK-n)$ is obtained from (\ref{scpp}) at $q\rightarrow 1$ and is equal to
\begin{equation}\label{npp}
A(N,N,\CK-n)=\prod_{k=1}^{N} \prod_{j=1}^N
\frac{\CK-n+j+k-1}{j+k-1}\,.
\end{equation}

A combinatorial description of the Schur functions may be given in terms of \textit{semi-standard Young tableaux}  \cite{fult}, which are in one-to-one correspondence with the nests of self-avoiding lattice paths.
A semi-standard Young tableau ${\sf{T}}$ of shape ${\bla}$ is a diagram whose cells are filled with positive integers $n\in \BN$ weakly increasing along rows and strictly increasing along columns.
Provided a semi-standard tableau ${\sf { T}}$ is given, the corresponding Schur function is defined as
\begin{equation}\label{eqsch}
S_{\bla}(x_1, x_2, \dots, x_m) =\sum_{\{{\sf { T}}\}} \textbf{x}^{\sf { T}}\,,
\qquad \textbf{x}^{\sf { T}} \equiv \prod_{i, j} x_{{\sf { T}}_{ij}}\,,
\end{equation}
where the monomial $\textbf{x}^{\sf {T}}$ is the \textit{weight} equal to the product over all entries ${\sf { T}}_{ij}$ ($i$ and $j$ label rows and columns of tableau ${\sf { T}}$). The sum in (\ref{eqsch}) is over all tableaux ${\sf { T}}$ of shape ${\bla}_N$ with the entries taken from the set $[m] \equiv \{1, 2, \ldots, m\}$, $m\geq N$.

\begin{figure}[h]
\centering
\includegraphics
{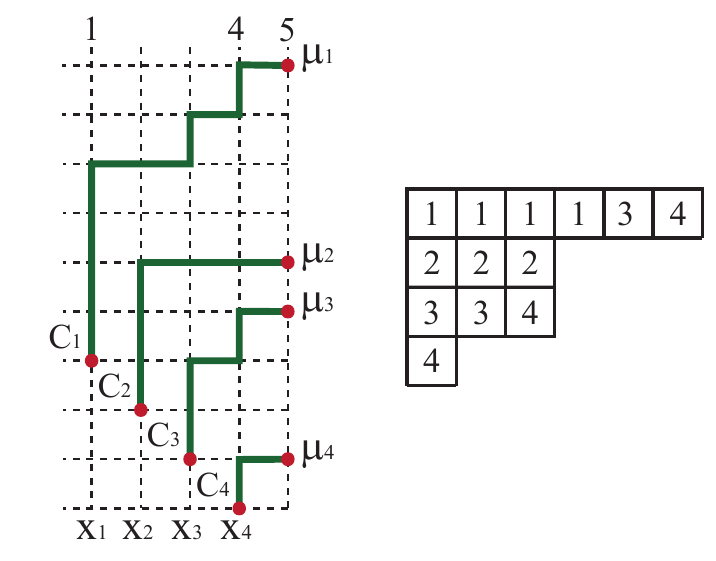}
\caption{A semistandard tableau of shape $\bla=(6, 3, 3, 1)$ as a nest $\cal{C}$ of lattice paths. The weight of $\cal{C}$ is $x_1^4 x_2^3 x_3^3 x_4^3$.}
\label{fig:f5}
\end{figure}

There is a natural way of representing each semi-standard tableau of shape $\bla$ with entries not exceeding $N$ as a nest of self-avoiding lattice paths with
prescribed start and end points. Let $T_{ij}$ be an entry in $i^{\rm th}$ row and $j^{\rm th}$ column of tableau $T$. The $i^{\rm th}$ lattice path (counted from the top of $T$)
encodes the $i^{\rm th}$ row of the tableau ($i=1, \ldots, N$). A nest $\cal{C}$ consists of paths going from points $C_i=(i, N-i)$ to points ($N, \mu_i=\la_i+N-i$) (see Fig.~\ref{fig:f5}).
Each path makes $\la_i$ steps to the north so that the steps along the line $x_j$ correspond to occurrences of the letter $j$ in the tableau $T$.
The power $l_j$ of $x_j$ in the weight of any particular nest of paths is the number of steps to north taken along the vertical line $x_j$. Thus, an equivalent representation of the Schur function is
\begin{equation}\label{schrepr}
S_{\bla} (x_1, x_2, \ldots, x_N) = \sum_{\{\cal{C}\}}\prod_{j=1}^{N}
x_{j}^{l_j},
\end{equation}
where summation is over all admissible nests $\cal{C}$.

From (\ref{schrepr}) it follows that the number of the described nests of paths is
\begin{equation}\label{numbpaths}
  S_{\bla} ({\bf 1}) = \sum_{\{\cal{C}\}}1=\prod_{1\leq j<k\leq K}\frac{\lambda_j - j - \lambda_k+k}{k-j}=\prod_{1\leq j<k\leq K}\frac{\mu_j-\mu_k}{k-j}\,,
\end{equation}
where $S_{\bla} ({\bf 1})\equiv S_{\bla} (1, 1, \ldots, 1)$,  \cite{macd}.

The $k^{\rm th}$ lattice path is contained within a rectangle of the size $\la_k\times (N-k)$, $1\le k \le N$. The starting point of each path is the lower left vertex. We define the {\it volume of the path} as the number of cells below the path within the corresponding rectangle. The volume of the nest of lattice paths $\cal C$ is equal to the volume of lattice paths:
\begin{equation}\label{weighta}
\mid \zeta \mid_{\cal{C}}\,\,= \sum_{j=1}^{N}(N-j)l_j= \sum_{j=1}^{N}(j-1)l_{N-j+1}.
\end{equation}
It follows that the $q$-parametrized Schur function is a partition function of the described nest:
\begin{equation}\label{schpar1}
S_{\bla}(\textbf{q})= \sum_{\{\cal{C}\}} q^{\mid \xi\mid_{\cal C}}=
   q^{\mid \bla \mid}\sum_{\{\cal{C}\}} q^{\mid \zeta \mid_{\cal C}}\,,
\end{equation}
where $|\bla|=\sum_{k=1}^N \la_k$ is the weight of partition.

\begin{figure}[h]
\center
\includegraphics {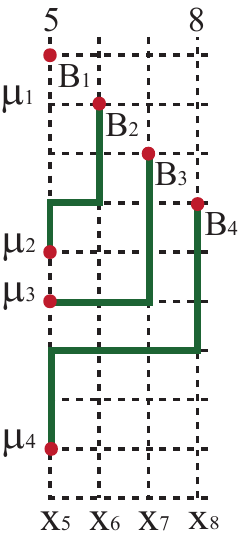}
\caption{Conjugated nest $\cal{B}$ of lattice paths.}
\label{fig:f6}
\end{figure}

The Schur function corresponding to the conjugate nest of self-avoiding paths is equal to
\begin{equation}\label{rsf}
S_{\bla} (y_1, y_2, \ldots, y_N) = \sum_{\{\cal{B}\}}\prod_{j=1}^{N}
y_{j}^{M-l_j},
\end{equation}
where summation is over all admissible nests $\cal{B}$ of $N$ self-avoiding lattice paths (see Fig.~\ref{fig:f6}).
The volume of the nest $\cal B$ is given by
\begin{equation}\label{weightab}
  \mid \zeta \mid_{\cal{B}}\,\,=\sum_{j=1}^{N} (j-1)(M-l_j)\,,
\end{equation}
and the partition function of $\cal B$ (see Fig.~\ref{fig:f6})
is obtained from (\ref{rsf}) in the parametrization ${\bf y}={\bf q}_N/q$:
\begin{equation}\label{schpar2}
S_{\bla} \Bigl(\frac{\textbf{q}_N}{q}\Bigr) = \sum_{\{{\cal B}\}} q^{\sum_{j=1}^{N} (j-1) ({\CM} - b_j)} \,,
\end{equation}
where Eq.~(\ref{weightab}) is used, and summation is over all admissible nests $\cal{B}$ of $N$ self-avoiding lattice paths.

\begin{figure}[h]
\center
\includegraphics {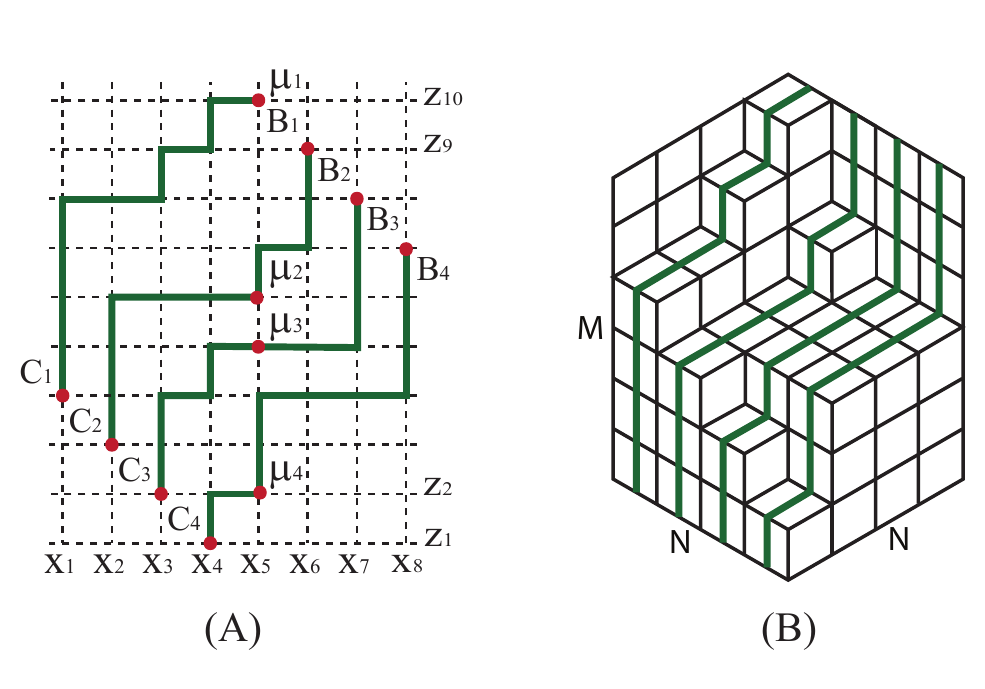}
\caption{Watermelon configuration and correspondent plane partition.}
\label{fig:f7}
\end{figure}

The scalar product, being the product of two Schur functions, may be graphically expressed as a nest of $N$ self-avoiding lattice paths starting at the equidistant points $C_i$ and terminating at the equidistant points $B_i$ ($1\le i \le N$). This configuration, known as a \textit{watermelon},  is presented in Fig.~\ref{fig:f7}. The scalar product is given by the sum of all such  watermelons. Rotating Fig.~7 by $\frac\pi 4$ counter-clockwise we see that the watermelon configuration is a particular case of configuration of paths for lock step random walkers (see Fig.~\ref{fig:f2}).

Using definitions (\ref{conwf1}) and (\ref{conj}) enables one to obtain the $q$-parameterized average of the projection operator $\bar\varPi_{n}$ (\ref{ratbe0}) (see (\ref{scqschr})):
\begin{eqnarray}
&& \langle \Psi_N ({\textbf{q}}^{-\frac{1}{2}})\mid \bar {\varPi}_n\mid \Psi_N ((\textbf{q}/q)^{\frac{1}{2}})\rangle \nonumber \\[0.2cm] && =
\sum_{\bla\backslash {\bf n} \subseteq
{\{(\CK-n)^N\}}}S_{\bla} (\textbf{q}) S_{\bla}(\textbf{q}/q)
=q^{nN^2} Z(N,N,\CK-n)  \,.\label{teproj}
\end{eqnarray}
The partition function of watermelons with the end points $C_i$, $B_i$, $1\le i\le N$ (the generating function of watermelons) is given by (\ref{teproj}) at $n=0$.

\section{Correlations Over $N$-Particles Ground State}

The transition amplitude
\begin{eqnarray}
&&\langle \Psi ({\textbf v}_N) \mid
\bar\varPi_{n}\, e^{-\frac{t}2 {\cal H}}\,\bar\varPi_{n}\mid\!\Psi ({\textbf u}_N)\rangle
\label{ampl0}\\
&& =\sum\limits_{{\bla^{L, R}\backslash {\bf n}}
\subseteq \{(\CK-n)^N\}}
S_{{\bla^L}}({\textbf v}^{-2}_N)\,S_{{\bla^R}}({\textbf u}^2_N)\,G({\bla^L}; {\bla^R}\,|\,t) \label{ampl1}\\
&& = \sum_{K=0}^{\infty} \frac{(t/2)^K}{K!}\sum \limits_{{\bla^{L, R} \backslash {\bf n}}
\subseteq \{(\CK-n)^N\}}
S_{{\bla^L}}({\textbf v}^{-2}_N)\,
S_{{\bla^R}}({\textbf u}^2_N)\,
\mathfrak{G}({\bmu^L}; {\bmu^R} | K)\, \label{ampl2}
\end{eqnarray}
is calculated with the help of Eqs. (\ref{conwf1}) and (\ref{conj}) \cite{bmumn}. In the above formulas
$\mathfrak{G}({\bmu^L}; {\bmu^R} | K)$ is given either by (\ref{qanal13}) or (\ref{ratbeco}), $\textbf u_N$ and $\textbf v_N$ stand for an arbitrary parametrization, and two independent summations over $\bla^{L, R} = {\bmu^{L, R}} - {\bdl}_N$ are analogous to those in (\ref{scschr}).

Substituting (\ref{ratbe7}) into (\ref{ampl1}), we obtain the transition amplitude in the form, \cite{bmnph}:
\begin{eqnarray}
\langle \Psi ({\textbf v}_N) \mid
\bar\varPi_{n}\, e^{-t {\cal H}}\,\bar\varPi_{n}\mid\!\Psi ({\textbf u}_N)\rangle
=\frac{1}{(M+1)^N
N!}\sum\limits_{s_1, \dots, s_N =0}^M
e^{t\sum\limits_{m=0}^N\cos \phi_{m}} && \nonumber  \\
\times\,|{\CV} (e^{i {\bphi}_N})|^2\, \mathcal{P}_{{\CK}-n}({\textbf{v}}^{-2},
e^{i\bphi})\, \mathcal{P}_{{\CK}-n}(e^{-i\bphi}, {\textbf{u}}^2)\,,&& \label{fa}
\end{eqnarray}
where $\mathcal{P}_{{\CK}-n}$ is the sum (\ref{scschr}).

Let us consider the expansion of the transition amplitude in the case when ${\bf u}_N^{2}={\bf v}_N^{2}={(1, 1, \ldots , 1)}$. From (\ref{ampl0})--(\ref{ampl2}) we obtain for the $K^{\rm th}$ term:
\begin{eqnarray}
&&\langle \Psi ({\textbf 1}) \mid
\bar\varPi_{n}\, (-{\cal H})^K\,\bar\varPi_{n}\mid\!\Psi ({\textbf 1})\rangle
\nonumber\\ [0.2cm]
&& = \sum \limits_{{\bla^{L, R} \backslash {\bf n}}
\subseteq \{(\CK-n)^N\}}
S_{{\bla^L}}({\textbf 1})\,
S_{{\bla^R}}({\textbf 1})\,
|P_K ({\bmu}^R \rightarrow\,{\bmu}^L)|\,, \label{ampl22}
\end{eqnarray}
where $|P_K ({\bmu}^R \rightarrow\,{\bmu}^L)|$ are integers given by (\ref{qanal27}),
and the values of the Schur functions $S_{{\bla^L}}({\textbf 1})$ and $S_{{\bla^R}}({\textbf 1})$ are given by (\ref{numbpaths}). If we put ${\bf u}_N^{2}={\bf v}_N^{2}={(1, 1, \ldots , 1)}$ in (\ref{fa}), a representation alternative to (\ref{ampl22}) is obtained:
\begin{eqnarray}
\langle \Psi ({\textbf 1}) \mid
\bar\varPi_{n}\, (-{\cal H})^K\,\bar\varPi_{n}\mid\!\Psi ({\textbf 1})\rangle\,=\,
\displaystyle{\frac{1}{(M+1)^N}}
\sum\limits_{\{{\bphi}_N\}}
\Bigr(2\sum\limits_{m=1}^N \cos {\phi}_{m}\Bigl)^K &&
\nonumber\\ [0.2cm]
\times\,|{\CV} (e^{i {\bphi}_N})|^2\, \mathcal{P}_{{\CK}-n}({\textbf{1}},
e^{i\bphi})\, \mathcal{P}_{{\CK}-n}(e^{-i\bphi}, {\textbf{1}}) \,. && \label{ampl32}
\end{eqnarray}
Since left-hand sides of (\ref{ampl22}) and (\ref{ampl32}) coincide, we obtain the following equality of two sums:
\begin{eqnarray}
\displaystyle{\frac{1}{(M+1)^N}}
\sum\limits_{\{{\bphi}_N\}}
\Bigr(2\sum\limits_{m=1}^N \cos {\phi}_{m}\Bigl)^K \,|{\CV} (e^{i {\bphi}_N})\, \mathcal{P}_{{\CK}-n}({\textbf{1}},
e^{i\bphi})|^2\, &&
\nonumber\\ [0.2cm]
= \sum \limits_{{\bla^{L, R} \backslash {\bf n}}
\subseteq \{(\CK-n)^N\}}
S_{{\bla^L}}({\textbf 1})\,
S_{{\bla^R}}({\textbf 1})\,
|P_K ({\bmu}^R \rightarrow\,{\bmu}^L)| \,. && \label{ampl42}
\end{eqnarray}
In other words, the trigonometric sum in left-hand side of (\ref{ampl42}) acquires an integer value expressed by the sum in right-hand side of (\ref{ampl42}).

\begin{figure}[h]
\center
\includegraphics [scale=1.1] {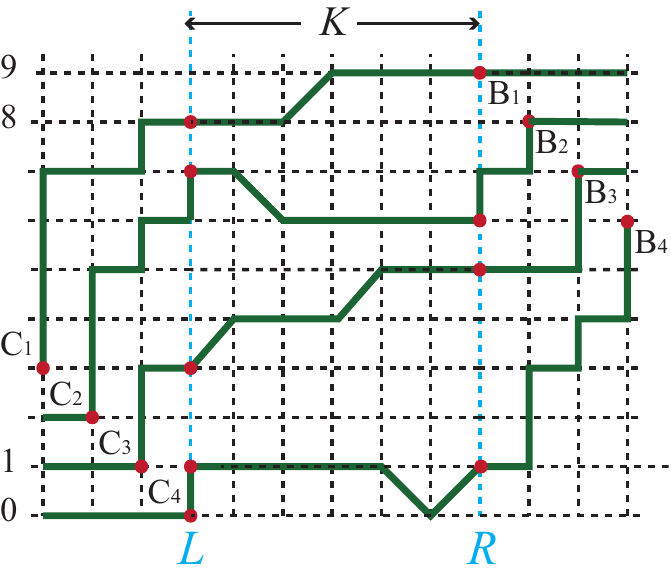}
\caption{Nest of paths contributing to $\langle \Psi ({\textbf 1}) \mid
\bar\varPi_{1}\,(-{\cal H})^K\,\bar\varPi_{1}\mid\!\Psi ({\textbf 1})\rangle$. }
\label{fig:f9}
\end{figure}

Summarising the graphical representations of functions involved in right-hand side of (\ref{ampl22}) we can give the  graphical representation
of the $K^{\rm th}$ term of the transition amplitude in terms of nests of self-avoiding lattice paths.
The first steps particles are doing
according the lock step rules starting from sites $C_i$ and finishing at accessible intermediate  positions $\bmu^R$. The number of
these nests is $S_{{\bla^R}}({\textbf 1})$. The next $K$ steps particles are doing according to the random turns rules
starting from sites $\bmu^L$ and terminating at $\bmu^R$. The number of these nests is $|P_K ({\bmu}^R \rightarrow\,{\bmu}^L)|$. The final steps are made again by the lock step rules from $\bmu^R$ up to $B_i$, and the number of these nests is $S_{{\bla^R}}({\textbf 1})$.
An example of the described nest of lattice paths is depicted in Fig.~\ref{fig:f9}.
Furthermore, two independent summations over intermediate positions $\bmu^{L, R}$ are meant in right-hand side of (\ref{ampl22}).
To obtain the transition amplitude we have to sum up over all intermediate steps $K$ (\ref{ampl2}).

Eventually, we use (\ref{fa}) and obtain the persistence of ferromagnetic string $\CT ({\bth}^{\,\rm g}_N, n,
t)$ (\ref{ratbe0}):
\begin{eqnarray} \label{ratbe131}
\CT ({\bth}^{\,\rm g}_N, n,
t)\,=\,\frac{1}{{\CN}^2({\bth}^{\,\rm g}_N) (M+1)^{N}}\,
\displaystyle{\sum\limits_{\{\bth_N\}} e^{-t (E_N({\bth}_N)-E_N(\bth^{\,\rm g}_N))}} && \nonumber \\
\times\,\bigl|\displaystyle{{\CV}_N(
e^{i{\bth}_N})\,
{\CP}_{\CK-n}(e^{-i{\bth}_N}, e^{i {\bth}^{\,\rm
g}_N})\bigr|^2}\,, &&
\end{eqnarray}
where summation is over all independent solutions of the Bethe equations (\ref{betheexp}). The sum
${\CP}_{\CK-n} (e^{-i{\bth}_N}, e^{i
{\bth}^{\,\rm g}_N})$ is given by (\ref{scschr}) on the solutions to Eqs.~(\ref{betheexp}), and  ${\CN}^2({\bth}^{\,\rm g}_N)$ is the squared norm of the ground state:
${\CN}^2({\bth}^{\,\rm g}_N)=\langle
\Psi ({\bth}^{\,\rm g}_N)
\mid\!\Psi ({\bth}^{\,\rm g}_N)\rangle$, where ${\bth}^{\,\rm g}_N$ are given by (\ref{grstxx}).

\section{Conclusion}

The representation of the Schur functions in terms of nests of self-avoiding lattice paths of lock step type, as well as the representation of the averages (\ref{qanal13}) in terms of random turns walks, made it possible for us to represent the correlation function of persistence of ferromagnetic string (\ref{ratbe0}) in the graphical way. Equation (\ref{ampl42}) is the main technical result of the paper obtained by means of the graphical approach developed.

\section*{Acknowledgement}

The research described
has been partially supported by RSF grant no.~16-11-10218.

\end{document}